\documentclass[shortnote,twocolumn]{jpsj3}
\usepackage{graphicx}

\title{Superconductivity in SrFe$_2$As$_2$ with Pt Doping}

\author{Yoshihiro \textsc{Nishikubo}$^{1,2}$, Satomi \textsc{Kakiya}$^{1,2}$, Masataka \textsc{Danura}$^{1,2}$, Kazutaka \textsc{Kudo}$^{1,2}$\thanks{E-mail: kudo@science.okayama-u.ac.jp} \\ and Minoru \textsc{Nohara}$^{1,2}$}
\inst{$^1$Department of Physics, Faculty of Science, Okayama University, 3-1-1 Tsushima-naka, Kita-ku, Okayama  700-8530, Japan \\ 
$^2$Transformative Research-Project on Iron Pnictides (TRIP), Japan Science and Technology Agency (JST), 5 Sanbancho, Chiyoda-ku, Tokyo 102-0075, Japan} 

\kword{iron-pnictide superconductor, SrFe$_2$As$_2$, electron doping, transition-metal substitution}

\begin{document}
\maketitle

The discovery of superconductivity with the superconducting transition temperature $T_{\rm c}$ of 26 K in LaFeAsO$_{1-x}$F$_x$ has triggered the intensive exploration of new Fe-based superconductors\cite{rf:Kamihara}. 
Competitive research has led to a rapid increase in $T_{\rm c}$\cite{rf:Ishida}, which reached the highest value to date of 56 K for Th-substituted GdFeAsO\cite{rf:CWang}. 
A variety of Fe-based superconductors have been discovered\cite{rf:Ishida}. 
Among them, AEFe$_2$As$_2$ (AE $=$ alkaline earth elements), abbreviated to 122, is fascinating because of its chemical versatility, namely, a wide variety of chemical elements can be substituted for AE, Fe and As to induce superconductivity. 
This gives us a chance to develop compounds with higher $T_{\rm c}$. 
Superconductivity appears in AEFe$_2$As$_2$ upon substituting alkali metals for AE (hole doping)\cite{rf:Rotter}, La for AE (electron doping)\cite{rf:Muraba}, transition metal elements TM such as Co\cite{rf:Jasper,rf:Chu} and Ni\cite{rf:Saha,rf:Li} for Fe (electron doping) and P for As (isovalent doping)\cite{rf:Ren,rf:Jiang,rf:Shi}. 
So far, the maximum $T_{\rm c}$ of 38 K has been observed in hole-doped Ba$_{1-x}$K$_x$Fe$_2$As$_2$\cite{rf:Rotter}. 
In contrast, electron-doped AEFe$_2$As$_2$ exhibits relatively low $T_{\rm c}$. 
SrFe$_2$As$_2$ exhibits superconductivity at $T_{\rm c} =$ 9.8 K and 8.7 K upon doping group 10 elements Ni\cite{rf:Saha} and Pd\cite{rf:Han}, respectively. 
In this paper, we report $T_{\rm c}$ = 17 K in electron-doped Sr(Fe$_{0.875}$Pt$_{0.125}$)$_2$As$_2$.

Polycrystalline samples of Sr(Fe$_{1-x}$Pt$_x$)$_2$As$_2$ ($x =$ 0.05, 0.075, 0.100, 0.125, 0.150, and 0.200) were synthesized by a solid-state reaction. 
FeAs precursor was first synthesized by heating Fe powder and As grains at 700$^\circ$C in an evacuated quartz tube. 
Then, prescribed amounts of Sr, FeAs, Pt and As powders or grains were mixed and ground carefully. 
The resulting powder was placed in an alumina crucible and sealed into an evacuated quartz tube. 
The ampule was heated at 700 $^\circ$C for 3 h and then at 1000 $^\circ$C for 24 h. 
After furnace cooling, the samples were ground, pelletized, wrapped with Ta foil and heated at 900$^\circ$C in an evacuated quartz tube. 
The products were characterized by powder X-ray diffraction. 
The samples were confirmed to be a single phase of the 122 structure. 
Electrical resistivity $\rho$ was measured by the standard DC four-terminal method in the temperature range between 2 and 300 K using Quantum Design PPMS. 
Magnetization $M$ was measured with a SQUID magnetometer (Quantum Design MPMS) from 5 to 25 K under a magnetic field of 10 Oe. 
\begin{figure}[b]
\begin{center}
\includegraphics[width=1\linewidth]{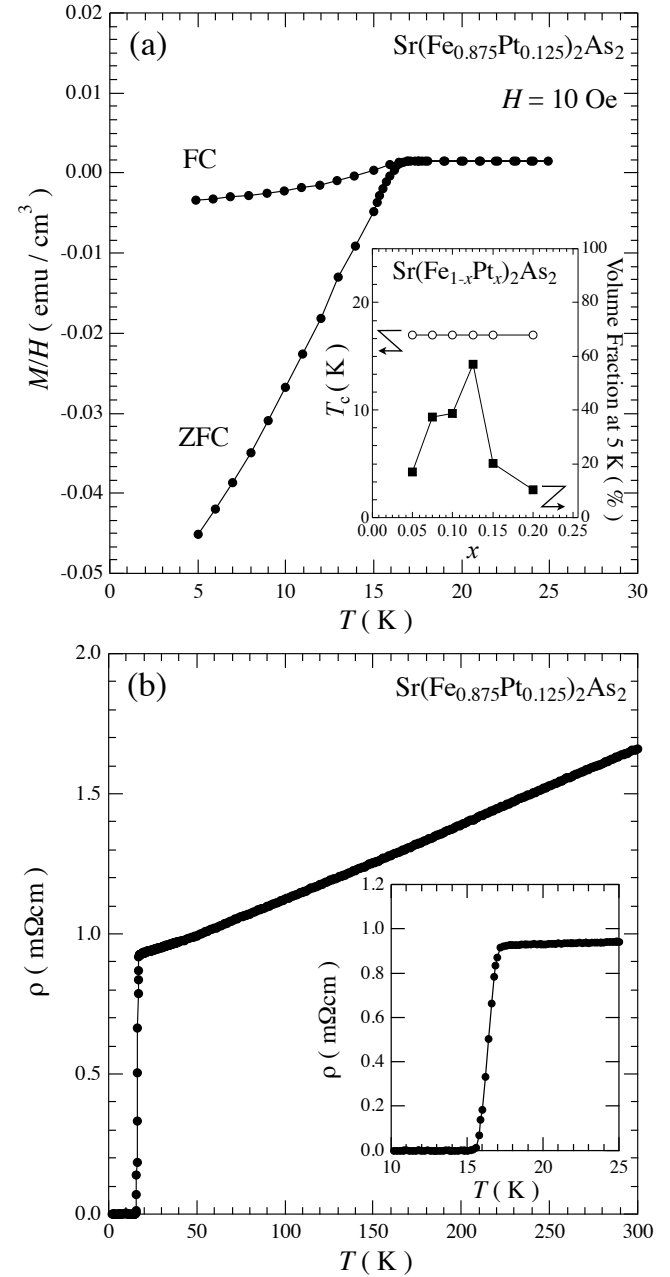}
\end{center}
\caption{(a) Temperature dependence of magnetization divided by applied field, $M/H$, of Sr(Fe$_{0.875}$Pt$_{0.125}$)$_2$As$_2$ at 10 Oe under zero-field-cooling (ZFC) and field-cooling (FC) conditions. 
The inset shows $x$ dependences of superconducting transition temperature $T_{\rm c}$ and the shielding volume fraction estimated at 5 K for Sr(Fe$_{1-x}$Pt$_{x}$)$_2$As$_2$. 
(b) Temperature dependence of electrical resistivity $\rho$ for Sr(Fe$_{0.875}$Pt$_{0.125}$)$_2$As$_2$. 
The inset is an enlarged figure in the vicinity of the superconducting transition. 
}
\label{f1}
\end{figure}

Figure 1(a) shows the temperature dependence of magnetization divided by applied field, $M/H$, of Sr(Fe$_{0.875}$Pt$_{0.125}$)$_2$As$_2$ at 10 Oe under zero-field-cooling and field-cooling conditions. 
$M$ exhibits diamagnetic behavior below about 17 K, indicating the occurrence of superconductivity at $T_{\rm c} =$ 17 K. 
The shielding volume fraction (VF) at 5 K is estimated to be 57 \% from the zero-field cooling data, supporting the existence of bulk superconductivity in Sr(Fe$_{0.875}$Pt$_{0.125}$)$_2$As$_2$. 
The inset shows $x$ dependences of $T_{\rm c}$ and VF. 
Superconductivity is observed in the range of 0.05 $\leq x \leq$ 0.200. 
VF exhibits a sharp peak at $x =$ 0.125, although $T_{\rm c}$ is almost independent of $x$, indicating the optimal doping at $x =$ 0.125 in Sr(Fe$_{1-x}$Pt$_{x}$)$_2$As$_2$.

Figure 1(b) shows the temperature dependence of $\rho$ for optimally doped Sr(Fe$_{0.875}$Pt$_{0.125}$)$_2$As$_2$. 
$\rho$ exhibits metallic behavior.
$\rho$ rapidly decreases below 17 K and zero resistivity is realized at 15.6 K, confirming the superconducting transition at $T_{\rm c} =$ 17 K. 
The parent compound SrFe$_2$As$_2$ exhibits long-range magnetic ordering and a structural phase transition at $T_0 =$ 198--220 K\cite{rf:Krellner,rf:Tegel,rf:Yan,rf:Zhao}, which manifest itself as a resistive anomaly\cite{rf:Jasper,rf:Saha,rf:Shi,rf:Han,rf:Krellner,rf:Yan}. 
For Sr(Fe$_{0.875}$Pt$_{0.125}$)$_2$As$_2$, in contrast, $\rho$ exhibits no anomaly above $T_{\rm c}$, indicating that the magnetic ordering and structural transition are completely suppressed by the Pt substitution of $x =$ 0.125 and the superconductivity occurs in the paramagnetic tetragonal phase.

\begin{table}[t]
\begin{center}
\caption{Maximum values of superconducting transition temperature $T_{\rm c}$ in transition-metal-substituted SrFe$_2$As$_2$ and BaFe$_2$As$_2$. AE and TM mean alkaline earth and transition metal elements, respectively. 
}\label{t1}
\begin{tabular}{|c|c|c|c|}
\hline
\makebox[10pt][c]{} & \makebox[45pt][c]{Group 8} & \makebox[45pt][c]{Group 9} & \makebox[50pt][c]{Group 10} \\
\hline
AE$\backslash$TM & Fe & Co & Ni \\
\hline
Sr  &       --       & 19.2 K\cite{rf:Jasper}  & 9.8 K\cite{rf:Saha}   \\
Ba &       --       & 24 K\cite{rf:Chu}          & 20.5 K\cite{rf:Li}  \\
\hline
       & Ru       & Rh        & Pd        \\
\hline
Sr  &  19.3 K\cite{rf:Schnelle}  & 21.9 K\cite{rf:Han} & 8.7 K\cite{rf:Han} \\
Ba &  20 K\cite{rf:Sharma}    & 23.2 K\cite{rf:Ni} & 19 K\cite{rf:Ni}  \\ 
\hline
       & Os       & Ir        & Pt        \\
\hline
Sr  &    --          & 24.2 K\cite{rf:Han} & 17 K [present work] \\
Ba &       --       & 28 K\cite{rf:Wang}  & 23 K\cite{rf:Saha2,rf:Zhu}  \\ 
\hline
\end{tabular}
\end{center}
\end{table}
Table I summarizes the maximum values of $T_{\rm c}$ for  SrFe$_2$As$_2$ and BaFe$_2$As$_2$  doped with various TM, together with that for the present Sr(Fe$_{0.875}$Pt$_{0.125}$)$_2$As$_2$. 
As can be seen from the table, robust superconductivity in 122 phases appears upon doping various TM elements, which range from group 8 to group 10 and from 3d to 5d. 
Isovalent Ru substitution leads to a moderately high $T_{\rm c}$. 
Group 9 and 10 elements act upon electron doping. 
Higher $T_{\rm c}$ is attained by doping group 9 elements than by doping group 10 elements. 
In addition, $T_{\rm c}$ tends to increase in the order of 3d, 4d, and 5d doping. 
In accord with this trend, Sr(Fe$_{0.875}$Pt$_{0.125}$)$_2$As$_2$ exhibits the highest $T_{\rm c}$ among SrFe$_2$As$_2$ doped with group 10 elements. 
These experimental observations suggest that the maximum $T_{\rm c}$ clearly depends on the substituted elements. 
At present, it is not clear which of the elements is the most favorable for attaining higher $T_{\rm c}$. 
However, it is now clear that Co, which has been the most-studied dopant so far, is not the best one. 
We propose that the substitution of various elements besides Co is important in enhancing $T_{\rm c}$ in iron-based superconductors.

In summary, we synthesized polycrystalline samples of Pt-substituted SrFe$_2$As$_2$ and measured the temperature dependence of magnetization and electrical resistivity. 
We observed the superconducting transition at $T_{\rm c} =$ 17 K with the maximum shielding volume fraction at $x =$ 0.125 in Sr(Fe$_{1-x}$Pt$_{x}$)$_2$As$_2$. 
It was found that the maximum $T_{\rm c}$ depends on the substituted element, so it is important to substitute various elements to explore new iron-based superconductors with higher $T_{\rm c}$.

\section*{Acknowledgment}
This work was partially performed at the Advanced Science Research Center, Okayama University.

\end{document}